# Ion-scale current structures in Short Large-Amplitude Magnetic Structures

Shan Wang[1,2*], Li-Jen Chen[2], Naoki Bessho[1,2], Michael Hesse[3,4], Lynn B. Wilson III[2], Richard Denton[5], Jonathan Ng[1,2], Barbara Giles[2], Roy Torbert[6], and James Burch[4]

[1]Astronomy Department, University of Maryland, College Park, MD 20742
[2]NASA Goddard Space Flight Center, Greenbelt, MD 20771
[3]Department of Physics and Technology, University of Bergen, Bergen, Norway
[4]Southwest Research Institute San Antonio, San Antonio, TX 78238
[5]Department of Physics and Astronomy, Dartmouth College, Hanover, NH 03755
[6]Space Science Center, University of New Hampshire, Durham, NH 03824
*swang90@umd.edu

Key points

- An ion inertial length ($d_i$) scale current sheet in SLAMS exhibits features of reconnecting current sheets

- Compression of SLAMS magnetic fields and magnetosonic whistler waves could generate $d_i$-scale current structures

- The current sheet and magnetosonic whistler waves produce comparable current densities and energy conversion rates

Abstract

We investigate electric current structures in Short Large-Amplitude Magnetic Structures (SLAMS) in the terrestrial ion foreshock region observed by the Magnetospheric Multiscale mission. The structures with intense currents ($|J| \sim 1\ \mu A/m^2$) have scale lengths comparable to the local ion inertial length ($d_i$). One current structure type is a current sheet due to the magnetic field rotation of the SLAMS, and a subset of these current sheets can exhibit reconnection features including the electron outflow jet and X-line-type magnetic topology. The $d_i$-scale current sheet near the edge of a SLAMS propagates much more slowly than the overall SLAMS, suggesting that it may result from compression. The current structures also exist as magnetosonic whistler waves with $f_{ci} < f < f_{lh}$, where $f_{ci}$ and $f_{lh}$ are the ion cyclotron frequency and the lower-hybrid frequency, respectively. The field rotations in the current sheets and whistler waves generate comparable $|J|$ and energy conversion rates. Electron heating is clearly observed in one whistler packet embedded in a larger-scale current sheet of the SLAMS, where the parallel electric field and the curvature drift opposite to the electric field energize electrons. The results give insight about the thin current structure generation and energy conversion at thin current structures in the shock transition region.

1. Introduction

How energy is converted from upstream bulk kinetic energy to downstream thermal and magnetic energies at collisionless shocks is a fundamental question of great interest. Poynting's theorem shows that the energy conversion between electromagnetic fields and particles occurs through $\bm{J} \cdot \bm{E}$, so currents within the shock transition region are naturally important for shock energy conversion.

It begs the questions of what the forms of current structures are, and what their relative importance in energy conversion is. Recent simulations [Karimabadi et al., 2014; Gingell et al., 2017; Bessho et al, 2019] and observations [Gingell et al., 2019a, 2019b; Wang et al., 2019] showed that some of the current sheets in the shock transition region can be reconnecting. Observations also showed that below 10 Hz magnetosonic whistler waves generate a significant fraction of the total current densities [Wilson III et al., 2014a, 2014b]. It would be valuable to compare the current density and energy conversion for the current sheets and whistler waves as well as for other possible forms of current structures.

Further, how the current structures are generated and evolve is an important question and not well understood. In our previous paper [Wang et al., 2019], the observed reconnecting current sheets are deep in the shock transition region: although the bulk ion speed is still decreasing, the magnetic field strength and plasma temperature are close to the downstream state, and continuous magnetic field fluctuations exist. However, a foreshock region with isolated Short Large-Scale Magnetic Structures (SLAMS), where the magnetic field strength is increased by more than twice of the ambient level [e.g., Schwartz et al., 1992] exists in that event. As the SLAMS evolve to become the new shock as suggested in observations [Schwartz et al., 1992] and simulations [Scholer et al., 2003; Tsubouchi et al., 2004]. In this study, we will use the same shock crossing as in Wang et al. [2019] to investigate whether SLAMS contribute to the generation of thin current structures, including reconnecting current sheets, and to examine the link between the foreshock and shock.

In the following, we will discuss three SLAMS, all containing current structures of ion inertial length ($d_i$) scales. The first contains a possibly reconnecting current sheet and magnetosonic whistler waves outside of the current sheet. The second contains current sheets that are being compressed, with magnetosonic whistler waves at its upstream edge. The third contains magnetosonic whistler waves with clear localized electron heating. The results demonstrate the association between SLAMS and current structures, and elucidate roles of thin current structures in energy conversion.

2. Data

The measurements are from the Magnetospheric Multiscale mission (MMS; Burch et al., 2016), during a crossing of the Earth's bow shock at GSE [8.4, 8.4, 0.1]$R_E$. Plasma data are from the Fast Plasma Investigation instrument (Pollock et al., 2016), with 150-ms resolution for ions and 30-ms resolution for electrons. Magnetic fields are from the fluxgate magnetometer (Russell et al., 2016) at 8 samples/s in the survey mode and 128 samples/s in the burst mode. Electric field data are from the axial (Ergun et al., 2016) and spin-plane double probes (Lindqvist et al., 2016) at 8,192 samples/s.

3. Results

Figure 1a shows the magnetic field in the foreshock region, where a series of pulsations exist and the SLAMS to be examined are denoted by arrows and numbers. The overview of the SLAMS 1 that we will examine (note it is not the first in time) is shown in the rest panels of Figure 1. The magnetic field (Figure 1b) is amplified with $B_{max}/|B_0|=4.9$, where $|B_0|=6.0$nT is the magnetic field strength in the upstream solar wind, and $B_{max}$ is the maximum magnetic field strength in the SLAMS. The density (Figure 1d) is enhanced with $n_{max}/n_0=3.9$, where $n_0=24$ cm$^{-3}$ is the upstream solar wind density and $n_{max}$ is the maximum density in the SLAMS. Inside the SLAMS, plasmas are decelerated and heated (Figures 1e-1g). Incident solar wind ions and reflected ions (possibly by the SLAMS) are deflected by the magnetic field in the SLAMS, as seen by the velocity variations of the two populations in the $V_x$ spectrogram (Figure 1e). The minimum variance direction (**k**) of the magnetic field during the interval between the vertical dashed lines is [0.974, 0.194, 0.116] GSE. The correlation analysis of the $B_z$ component of the magnetic field measured by four spacecraft during 13:24:36.5-13:24:39.5 UT suggests the propagation direction to be [0.997, -0.072, -0.012] GSE, roughly consistent with the minimum variance direction with a difference of 17 degrees. The spacecraft frame propagation speed of the SLAMS is -155 km/s, i.e., anti-sunward. The upstream solar wind speed is determined by looking for the centroid of contours in the distribution [Wilson III et al., 2014a] during 13:20:10-13:20:12 UT, which is 342 km/s roughly along GSE -x direction. Thus, in the upstream solar wind frame, the propagation is 187 km/s sunward, corresponding to 6.9 $V_A$, where $V_A=27$km/s is the upstream Alfvén speed. The propagation speed is close to while greater than that in a previous statistical study of SLAMS of 1-6 $V_A$ [Mann et al., 1994]. Figure 1j shows the hodogram of the magnetic field in the $B_i$-$B_j$ plane for the marked interval, where i and j represent the maximum and intermediate variance directions, respectively. The k component of the magnetic field at upstream is negative (out of the page, as seen in $B_x$<0 in Figure 1a). The counter-clockwise rotation of the magnetic field from red to blue indicates right-hand polarization around the magnetic field in the spacecraft frame, and hence left-hand polarization in the solar wind frame. The scale of the SLAMS during 8 s is 1240 km ~ 26 $d_{i0}$, where 1$d_{i0}$=47km is the upstream ion inertial length. The ~1000 km scale of the SLAMS is consistent with previous observations [Lucek et al., 2008].

An intense current sheet (Figure 1h) with reconnection features is observed in the middle of this SLAMS. The magnetic field has a sharp rotation at ~13:24:40 UT with reversals of $B_y$ and $B_z$. The rotation is part of the SLAMS, which is during the counter-clockwise rotation in the upper right quadrant in the $B_i$-$B_j$ hodogram in Figure 1j (marked by the black arrows). The rotation is left-handed in the upstream solar wind frame. Near the end of the hodogram, the light-to-dark blue trace in the upper left quadrant exhibits clockwise loops, corresponding to the magnetic field variations at 13:24:40-13:24:43 UT outside of the current sheet. This part of the magnetic field variation is the magnetosonic whistler wave with right-handed polarization in the plasma rest frame (defined by the local ion bulk velocity including all ion components).

The current sheet is possibly reconnecting as suggested by the electron outflow jet. Figure 2 shows the zoom-in view of the current sheet, where the vectors are rotated to the LMN coordinate determined by the minimum variance analysis (MVA) across the current sheet (see Figure 2 caption for the transformation matrix between GSE and LMN). The sharp $B_L$ reversal is associated with negative $V_{eM}$ enhancements. The electron bulk flow (Figure 2b) exhibits a positive peak of $V_{eL}$=150km/s (2.8$V_{A,loc}$), where $V_{A,loc}$=54km/s is the average Alfvén speed across the current sheet during 13:24:38.5-13:24:41.0 UT, while the $V_{eL}$ outside of the current sheet (at the edges of the shown interval) is near-zero. We note that $B_M$ has quadrupolar variations across the current sheet instead of the bipolar Hall fields as in standard reconnecting current sheets, possibly because higher-frequency waves are superimposed on the current sheet. The $V_{eL}$ peak near 13:24:39.8 UT is associated with the $B_M$ rise. The propagation speeds determined by the correlation analysis of $B_L$ and $B_M$ are close to each other within 10 km/s, and hence the $B_M$ (as well as the $V_{eL}$) variation is considered to be part of the current sheet with reversing $B_L$.

The current sheet convection speed based on the correlation analysis during 13:24:39.6-13:24:40.1 UT is 144 km/s in the spacecraft frame, close to the propagation speed of the SLAMS (155 km/s). The corresponding current sheet thickness is 2.3 $d_{i0}$ and 1.5 $d_{i,loc}$, where $d_{i,loc}$=31 km is based on the average n = 52 cm$^{-3}$ across the current sheet. As discussed in Figure 1e, individual populations of incoming solar wind and SLAMS reflected ions are deflected by the magnetic field, resulting in velocity variations in the spectrogram over a much larger scale than the current sheet. However, the L component of bulk ion velocity has little variation within the current sheet (Figure 2c), i.e., no ion outflow jet is formed. The resulting current density (Figure 2d) is dominated by the parallel component, reaching $1.3 \mu A/m^2$, and $\boldsymbol{J} \cdot \boldsymbol{E'} = \boldsymbol{J} \cdot (\boldsymbol{E} + \boldsymbol{V_e} \times \boldsymbol{B})$ (Figure 2e) is enhanced at the peak $V_{eL}$ jet. The electron temperature (Figure 2f) is higher in the central region of the SLAMS (earlier in time, also seen in Figure 1g) and fluctuates along with the magnetic field strength at the magnetosonic wave during 13:24:40-13:24:43 UT, but does not show particular enhancements inside the current sheet, i.e., no clear heating directly associated with the thin current sheet.

The possibility of reconnection is further supported by reconstructed magnetic field structures using four-spacecraft magnetic field and plasma current density measurements (plasma current densities are interpolated to the magnetic field time cadence). We employ the reconstruction based on the 2$^{nd}$-order polynomial expansion relative to the fields at the spacecraft barycenter [Torbert et al., 2019; Denton et al., 2020]:

$$B_L \sim [B_{L0}] + \left[\frac{\partial B_L}{\partial L}\right]L + \left[\frac{\partial B_L}{\partial M}\right]M + \left[\frac{\partial B_L}{\partial N}\right]N + \left[\frac{\partial^2 B_L}{\partial N^2}\right]N^2 \quad (1)$$

$$B_M \sim [B_{M0}] + \left[\frac{\partial B_M}{\partial L}\right]L + \left[\frac{\partial B_M}{\partial M}\right]M + \left[\frac{\partial B_M}{\partial N}\right]N + \left[\frac{\partial^2 B_M}{\partial L^2}\right]L^2 + \left[\frac{\partial^2 B_M}{\partial N^2}\right]N^2 + \left[\frac{\partial^2 B_M}{\partial L \partial N}\right]LN \quad (2)$$

$$B_N \sim [B_{N0}] + \left[\frac{\partial B_N}{\partial L}\right]L + \left[\frac{\partial B_N}{\partial M}\right]M + \left[\frac{\partial B_N}{\partial N}\right]N + \left[\frac{\partial^2 B_N}{\partial L^2}\right]L^2 \quad (3)$$

The terms in the brackets are 17 unknowns, including the magnetic field at the barycenter ($B_{L0}$, $B_{M0}$, $B_{N0}$), and the magnetic field gradients. A global LMN coordinate determined by MVA is used. The above expansion is the 'reduced 2$^{nd}$-order' form, which includes a few 2$^{nd}$-order terms that are expected to be important for a reconnection-like current sheet with the gradients mainly in the L-N plane, while neglecting terms that are expected to be small ($\frac{\partial^2 B_L}{\partial L^2}$, $\frac{\partial^2 B_N}{\partial N^2}$, $\frac{\partial^2 B_j}{\partial M \partial i}$, $\frac{\partial^2 B_k}{\partial L \partial N}$ where i, j=L, M or N, and k=L or N) [Denton et al., 2020]. Equations (1)-(3) can be evaluated at the individual spacecraft positions. Along with $\nabla \times B = \mu_0 j$

(for three components) and $\nabla \cdot B = 0$, we have 25 equations in total, and the unknowns could be solved through the least mean square method.

Figure 3 shows the reconstruction result. During the current sheet crossing at 13:24:39.77-13:24:40.00 UT, the reconstruction gives small $|\nabla \cdot B|/|\nabla \times B|$ (Figure 3b, less than 10%), nearly identical magnetic fields between reconstruction and measurements (not shown), and the good agreement between the reconstructed (dashed) and measured (solid) current densities (Figures 3c-3f), which indicates good reconstruction results for this interval. The reconstructed magnetic fields produce an X-line topology in the L-N plane at M=0 (barycenter) during the two marked intervals (13:24:39.782-13:24:39.813 UT and 13:24:39.884-13:24:39.930 UT). An example at the end of the 1$^{st}$ interval during the $V_{eL}$ jet is shown in the bottom panel. An X-line exists at an L distance of ~20 km (0.64 $d_{i,loc}$, 2.3 $L_{sc}$, where $L_{sc}$=8.7 km is the average inter-spacecraft separation) from the spacecraft barycenter. The plots of magnetic field lines in these two intervals are shown in Figures S1 and S2 of the supplementary information. In these two intervals, the location of the X-line varies in a way consistent with the spacecraft passing from the -N to +N side of the current sheet. In addition, an X-line could also be reproduced if using the local LMN coordinate based on the MDD method [Shi et al., 2005] to perform the polynomial expansion [Denton et al., 2019], and the linear polynomial expansion [Fu et al., 2015] (Figure S3). Although the structures of the current sheet from various methods and intervals of the reconstruction are not identical, the existence of the X-line is robustly suggested by reconstruction, supporting the current sheet to be reconnecting.

The SLAMS event 2 (marked in Figure 1a) is shown in Figure 4. It has $|B_{max}|/|B_0|$=6.9, and $n_{max}/n_0$=4.3. The propagation in the spacecraft frame from the correlation analysis of $B_z$ measured by four spacecraft during the reversal at 13:24:58.5-13:25:02.0 UT is -148×[0.936, 0.350, -0.042] GSE km/s. In the upstream solar wind frame, the SLAMS propagates toward upstream with a speed of 194 km/s (7.3 $V_A$), with left-hand polarization. The scale size of the SLAMS along the propagation direction during 13:24:56.0-13:25:04.5UT is 1258 km (27 $d_{i0}$).

Current structures of the $d_i$ scale exist at the edges of the SLAMS. The downstream edge of the SLAMS (earlier in time) has reversals of $B_y$ and $B_z$ during 13:24:56.0-13:24:56.5 UT (marked as cs I), with a current density up to 1.6 $\mu A/m^2$ (Figure 4g). The propagation velocity in the spacecraft frame is 91×[-0.898, -0.440, 0.000] GSE km/s, and the scale is 1 $d_{i0}$. Near the density gradient, there is a sharp $B_y$ rise during 13:24:57.5-13:24:59.3 UT (marked as cs II), with a current density enhancement of ~1.03$\mu A/m^2$ (Figure 4g). The propagation velocity is 65×[-0.935, -0.339, 0.108] GSE km/s, and the scale length of the $B_y$ rise is 2.5 $d_{i0}$. The propagation speeds of both $d_i$-scale current sheets are much smaller than overall propagation speed of the SLAMS, suggesting that the downstream edge of the SLAMS is being compressed, which might contribute to generating the thin current sheets.

The upstream edge of the SLAMS (13:25:02-13:25:04.5 UT) has magnetosonic whistler wave fluctuations, a common feature of steepening SLAMS [e.g., Schwartz et al., 1992, Wilson III et al., 2013], locally generating current densities up to 1.63 $\mu A/m^2$ and enhancements of $\boldsymbol{J} \cdot \boldsymbol{E'}$ (Figure 4h). In the spacecraft frame, the wave frequency is 1.5 Hz, and the phase speed ($V_{ph}$) obtained from the correlation analysis of magnetic fields is 87 km/s, propagating at 34 ° from the quasi-steady magnetic field and 37 ° from the propagation of the SLAMS. The fluctuations are right-handed in the plasma rest frame (blue clockwise loops in the $B_i$-$B_j$ plane of the hodogram in Figure 4i). Thus, we are observing magnetosonic whistler waves. The whistler waves have the corresponding $kd_i$=3.9, where $d_i$=34 km is based on the average density during 13:25:02-13:25:04.5 UT. During the whistler interval, the ion bulk velocity (including both incoming solar wind and shock/SLAMS reflected ions) along **k** is 274 km/s. Thus, in the plasma rest frame, $V_{ph}$=187 km/s, f=3.2Hz=0.34 $f_{lh}$, where $f_{lh} = \sqrt{f_{ci}f_{ce}}$=9.3Hz is the lower hybrid frequency. The magnetic field and electron bulk flow oscillate together, without a jet signature that breaks the correlation between the two as in traditional reconnection events. However, we do not rule out the possibility that reconnection will occur inside the whistler wave packets, since the associated thin current structures provide a necessary condition for reconnection. Compared to the downstream edge with a $B_y$ rise, the upstream edge with decreasing $B_y$ is less steep. Its spacecraft frame propagation speed determined at 13:24:02 UT is 121 km/s, slower than the overall SLAMS and faster than the whistler wave. The $d_i$ scale whistler wave grows on top of the larger-scale SLAMS edge that is steepening. The electron temperature has visible fluctuations in the parallel and perpendicular components (Figure 4f), but no substantial net heating is observed at the magnetosonic whistler waves or current sheets I and II in this second SLAMS event.

The SLAMS event 3 is shown in Figure 5. Considering the whole structure as one SLAMS, it has $|B_{max}|/|B_0|$=3.1, $n_{max}/n_0$=2.3. High-frequency fluctuations exist in the middle of the

SLAMS. The propagation in the spacecraft frame from the correlation analysis of <0.5 Hz $B_y$ during the reversal at 13:23:57-13:24:01 UT is -106×[0.910, 0.247, -0.333] GSE km/s. In the upstream solar wind frame, the SLAMS propagates sunward with a speed of 236 km/s (8.7 $V_A$), with left-hand polarization (overall counter-clockwise rotation from red to blue in Figure 5i). This SLAMS has gradual gradients at both edges. Taking the marked interval of 13:23:51-13:24:03 UT, the scale size of the SLAMS along the propagation direction is 1272 km (27 $d_{i0}$).

The high-frequency magnetosonic whistler waves in the middle of the SLAMS lead to a current density up to 1.1 $\mu A/m^2$ (Figure 5g) and enhancements of $\boldsymbol{J} \cdot \boldsymbol{E}'$ (Figure 5h). The spacecraft-frame frequency of the wave is ~2.0 Hz, and the wave propagates anti-sunward. In the local plasma rest frame, $V_{ph}$=133 km/s sunward, 22° from the quasi-steady magnetic field (<0.5 Hz), f=2.0Hz=0.26$f_{lh}$, k$d_i$=3.4. Overall, $T_{e\perp}$ (Figure 5f) increases toward the SLAMS center as the magnetic field strength increases. In the magnetosonic whistler wave interval, a dramatic $T_{e\parallel}$ peak comparable to the net perpendicular heating into the SLAMS appears, associated with a parallel electron beam in the distribution (Figure 5j).

The electron energization around the $T_{e\parallel}$ peak is further analyzed in Figure 6. The SLAMS structure is associated with a current sheet with magnetic field reversal in GSE $B_y$ (Figure 5b), while the magnetosonic whistler waves lead to sharper variations of the magnetic field. The magnetic field is rotated to the LMN coordinate determined by MVA of 1-5Hz fields during 13:23:58.28-13:23:58.54 (Figure 6a), a coordinate that gives a clear reversal of $B_L$ and the electron curvature drift velocity using four-spacecraft measurements [Shen et al.,

2003] mainly along the out-of-plane -M direction (Figure 6f). The magnetic field strength becomes low in the middle of the current sheet (black curve in Figure 6a). For electrons that can be trapped in the current sheet and mirrored at the edge of the central current sheet where |B|$_{max}$ is 17 nT, their pitch angles $\alpha = \mathrm{asin}\left(\frac{|B|}{|B|_{max}}\right)$ [Lavraud et al., 2016] are shown as black curves on top of the pitch angle distribution of 15-60 eV electrons, the energy range with clear energization as seen in the omni-directional spectrogram (Figure 6b). The lower magnetic field strength in the current sheet center leads to more field-aligned pitch angle distributions, which contribute to the increase of T$_{e\parallel}$. On the other hand, the total energy is increased (Figure 6b), demonstrating net energization in addition to the effect of the mirror force.

Both parallel and perpendicular electric fields contribute to the electron energization, as shown in $\boldsymbol{J_e} \cdot \boldsymbol{E}$ (Figure 6d), where $\boldsymbol{J_e}$=-ne$\boldsymbol{V_e}$ measured by MMS1, and $\boldsymbol{E}$ is the electric field, both are transformed to the local current sheet frame with a motion of V$_{cs}$=-146 km/s along N determined by the four-spacecraft magnetic field correlation analysis. Electrons in the parallel beam in Figure 5j are most clearly energized, which have a parallel velocity of about 2500 km/s and an energy of 18 eV. It is at the time marked by the first vertical dashed line in Figure 6. Figure 6e shows the 1D electron distribution along $U_\parallel = sign(V_\parallel)\frac{1}{2}m_e V_\parallel^2$ cut at the bulk perpendicular velocity. V$_\parallel$>0 electrons move from the B$_L$>0 side toward the B$_L$<0 side (from the right to the left side of the plot). The distribution at $U_\parallel > 15\ eV$ at the first vertical line is elevated by one bin (3 eV) compared to that at the second vertical line, indicating that electrons are energized by 3 eV as they move in the N direction from the second to the first vertical line. Since the parallel beam has the same energy of 18 eV with

$T_{e\parallel}$, $V_{curv}$ calculated using $T_{e\parallel}$ (Figure 6f) represents the curvature drift for the parallel beam. The energy conversion rate $-eV_{curv} \cdot E$ fluctuates with positive and negative values. During dt=0.09 s between the two vertical dashed lines, the N distance is $\Delta N = V_{cs}dt$ =13.1 km. The magnetic field is close to 45° from the N direction, so the duration for an electron with V$_\parallel$=2500 km/s to move across $\Delta N$ is $\Delta t = \Delta N/(V_\parallel \sin(45°))$~0.0073 s. $-eV_{curv} \cdot E$ is about 50 eV/s, so that the electron energy gain is $-eV_{curv} \cdot E \cdot \Delta t$~0.4 $eV$. The parallel electric field (red curve in Figure 6h with burst mode resolution) close to -1 mV/m near the current sheet center is barely more significant than the estimated uncertainty (blue shade). We estimate the energization by the parallel electric field between the two vertical dashed lines using $\int -eV_\parallel E_\parallel dt$ to be 3.6 eV, where $-eV_\parallel E_\parallel$ averaged to the electron velocity time cadence is shown in Figure 6i, though the number needs to be taken with cautions since not all data points of E$_\parallel$ in the interval have larger amplitudes than the uncertainty. Based on the above estimation, the energization by E$_\parallel$ and the curvature drift opposite to the electric field for the parallel drifting electrons is about 4.0 eV, close to the observed energization of 3 eV.

4. Summary and discussions

In this study, we investigate the current structures in the Earth's foreshock region, in SLAMS in particular. The most intense current structures with the current density of ~1 $\mu A/m^2$ are of the d$_i$ scale, and are associated with energy conversion $\boldsymbol{J} \cdot \boldsymbol{E}'$.

The current structures could be in the form of current sheets that are part of the magnetic field rotation in SLAMS (in the 1$^{st}$ and 2$^{nd}$ SLAMS discussed above), which are possibly reconnecting (in the 1st SLAMS) as suggested by the electron outflow jet and reconstructed X-line-like magnetic field structures. They are also observed in the form of magnetosonic whistler waves with the rest-frame frequency $f_{ci} < f < f_{lh}$ (in the 2nd and 3rd SLAMS), which are generated superimposed on the SLAMS structure. The two forms of the current structures have comparable current density and $\boldsymbol{J} \cdot \boldsymbol{E'}$ values, and fluctuations of the electron flows have similar amplitudes.

The reconnecting current sheet in SLAMS 1 discussed above is part of the magnetic field rotation in the SLAMS. It suggests that the thin reconnecting current sheet evolves in association with the compression of the SLAMS. The compression is indeed observed near the edge of the second SLAMS, where the spacecraft-frame propagation speed at the sharp magnetic field gradient is less than half of that of the overall SLAMS determined from the gradual magnetic field rotation in the middle. The low magnetic field strength in the current sheet that is clearest in current sheet I of the 2nd SLAMS is a favorable condition for the compression. It is possible that with further compression, the current sheet (already with a scale of only 2.5 $d_i$) could further thin down and may reconnect.

Parallel electron heating associated with a parallel beam is observed simultaneously with the magnetosonic whistler wave in the 3rd SLAMS. In previously reported reconnection events [Gingell et al., 2019a; Wang et al., 2019], the reconnecting current sheets with only electron jets do not exhibit net electron heating, while a current sheet with ion jets show ion and electron heating. These observations suggest that both the current sheets and waves cause plasma heating, but not always. The analysis of the 3rd SLAMS indicates that the small-scale magnetosonic whistler wave superimposed on the larger-scale current sheet enhances the magnetic field curvature and produce parallel electric fields in the current sheet, which possibly enhances electron energization. For the reconnecting current sheet in the 1$^{st}$ SLAMS, the $B_M$ variations that differ from the standard Hall field structures are also likely the signatures of high-frequency waves superimposed on the current sheet. These observations suggest that the coupling of multiple-scale current structures may result in more efficient electron energization.

The results in this study suggest that the foreshock structures like SLAMS provide initial locations and magnetic field fluctuations to generate thin current structures. The SLAMS and current sheets are then propagated to the shock, while more current structures are generated. Further investigations with observations and simulations will help in understanding the entire process of generation and evolution of the $d_i$-scale waves and current sheets, and quantifying their roles in the energy conversion at shocks.

Acknowledgments

The research at UMCP and GSFC is supported in part by DOE grant DESC0016278, DESC0020058, NSF grants AGS-1619584, NASA 80NSSC18K1369, and the NASA MMS mission. Work at Dartmouth College is supported by NASA 80NSSC19K0254. The work is also supported by the International Space Science Institute's (ISSI) International Teams programme. MMS data are available at MMS Science Data Center (https://lasp.colorado.edu/mms/sdc/).

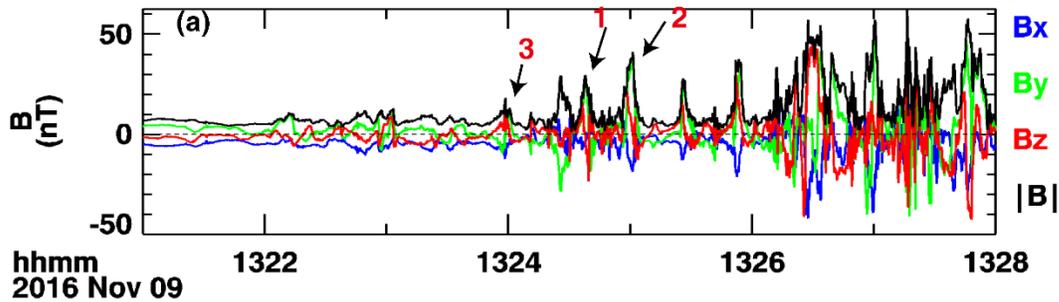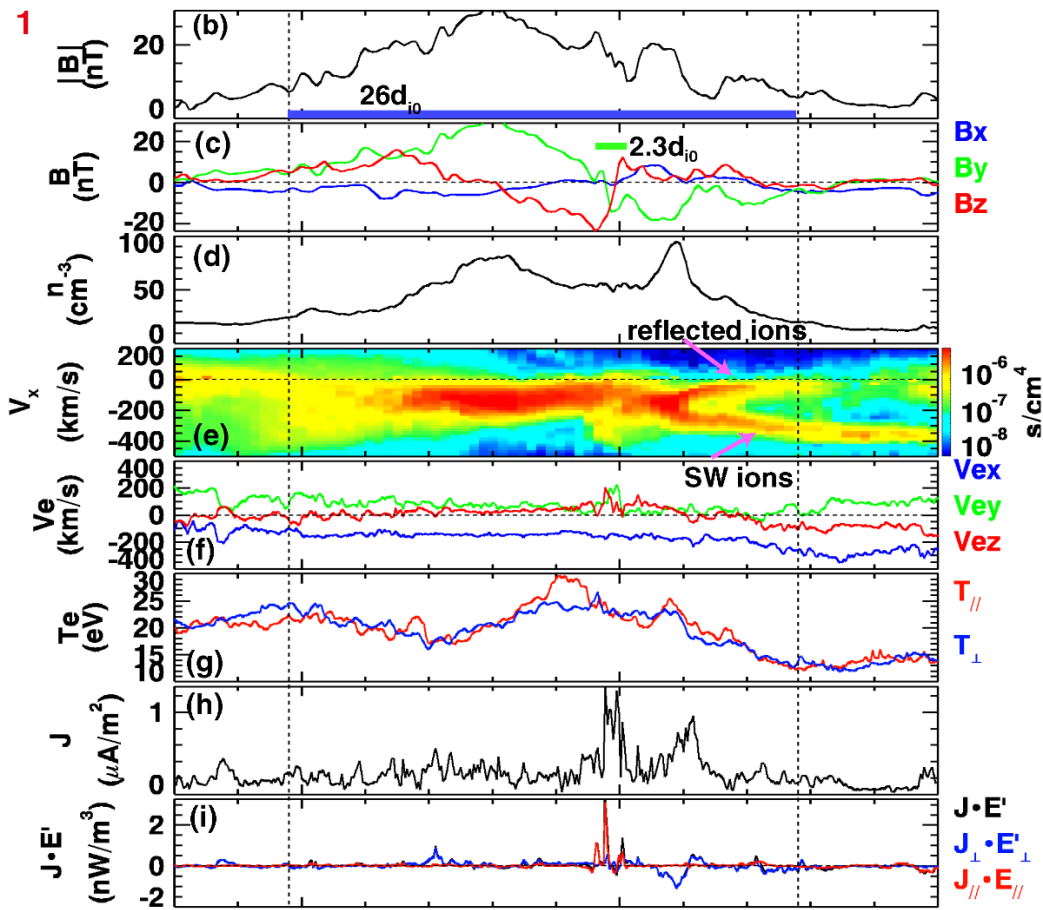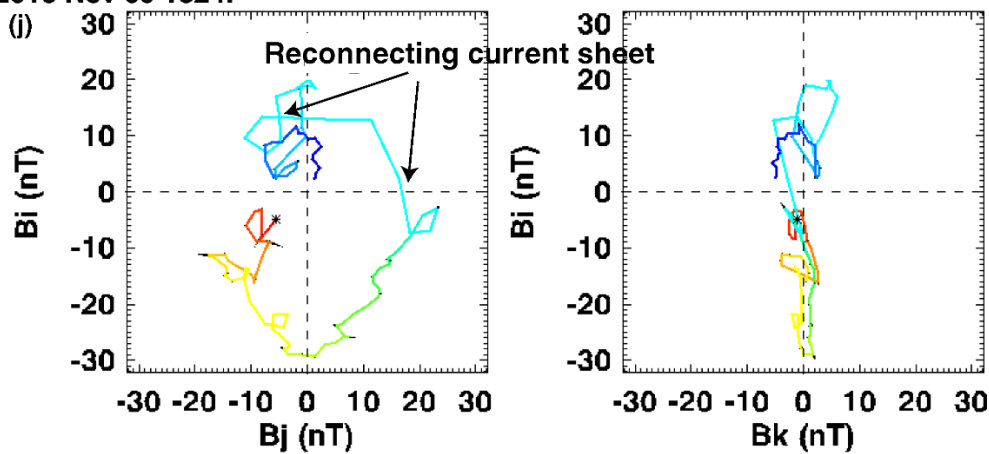

**Figure 1.** (a) Magnetic field in the foreshock region with isolated pulses, measured by MMS1. The three SLAMS discussed in the paper are marked. (b)-(i) overview of the SLAMS event 1. (b) magnetic field strength; (c) magnetic field vector in GSE (d) electron density; (e) ion spectrogram along GSE $V_x$; (f) electron velocity; (g) electron temperature; (h) current density; (i) electron frame energy conversion rate. (j) hodogram of the magnetic field during the interval marked by the dashed vertical lines in (b)-(i), where i, j, and k represent the maximum, intermediate, and minimum variance directions, respectively. A $d_i$-scale current sheet as part of the magnetic field rotation exists around 13:24:40 UT, possibly reconnecting as demonstrated in Figures 2-3.

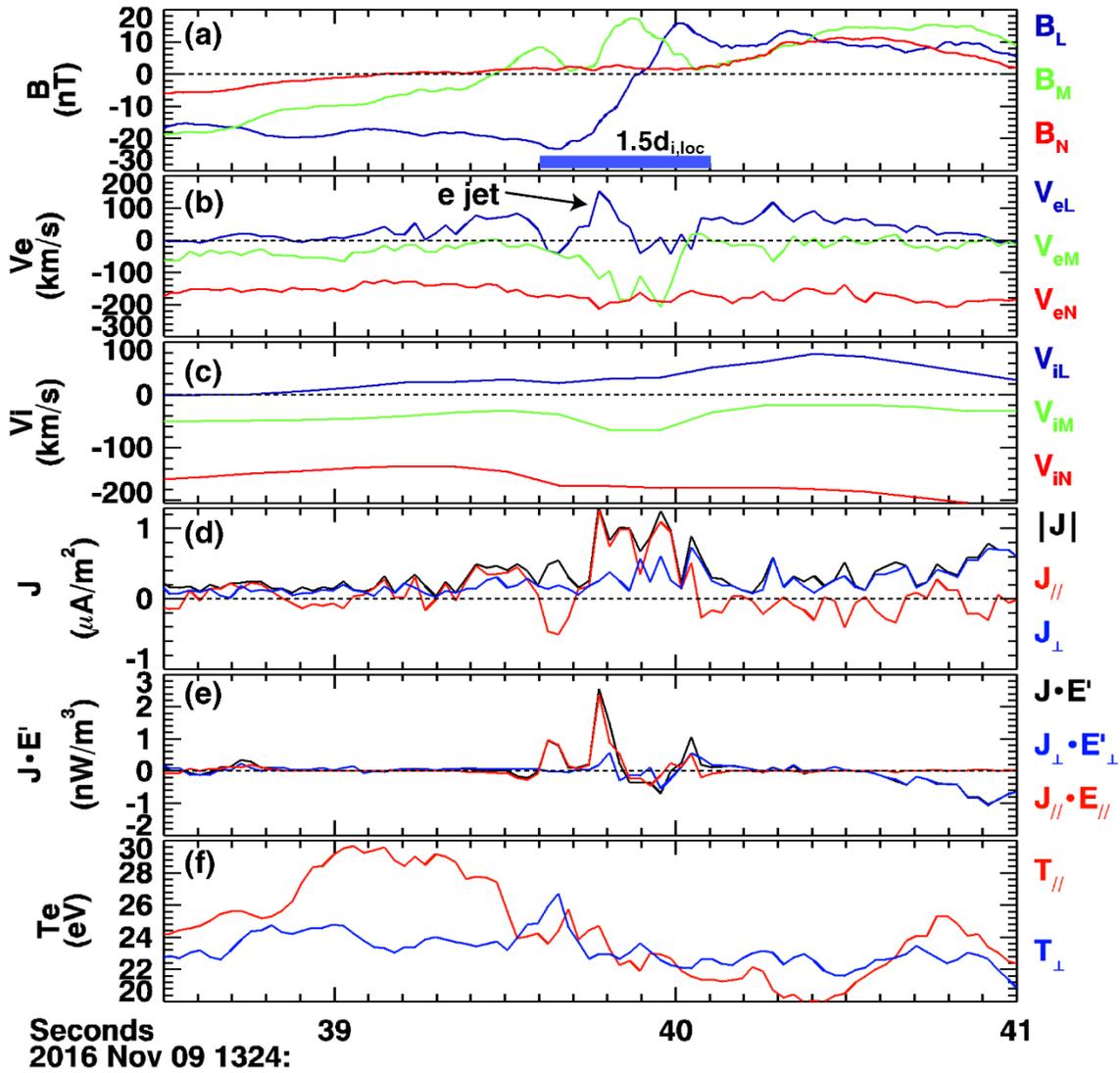

**Figure 2.** Reconnecting current sheet in the first SLAMS. (a)-(c) magnetic field, electron and ion velocities in the LMN coordinate. The LMN coordinate is determined using MVA during 13:24:39.4-13:24:40.2 UT, where L=[0.0322, -0.4376, 0.8986], M=[-0.1933, -0.8488, -0.4240], N=[0.9806, -0.1601, -0.1131] GSE. During the increase of $B_L$, a peak $V_{eL}$ jet occurs, associated with parallel current density (d) and energy conversion (e). The electron temperature does not exhibit enhancements in the current sheet.

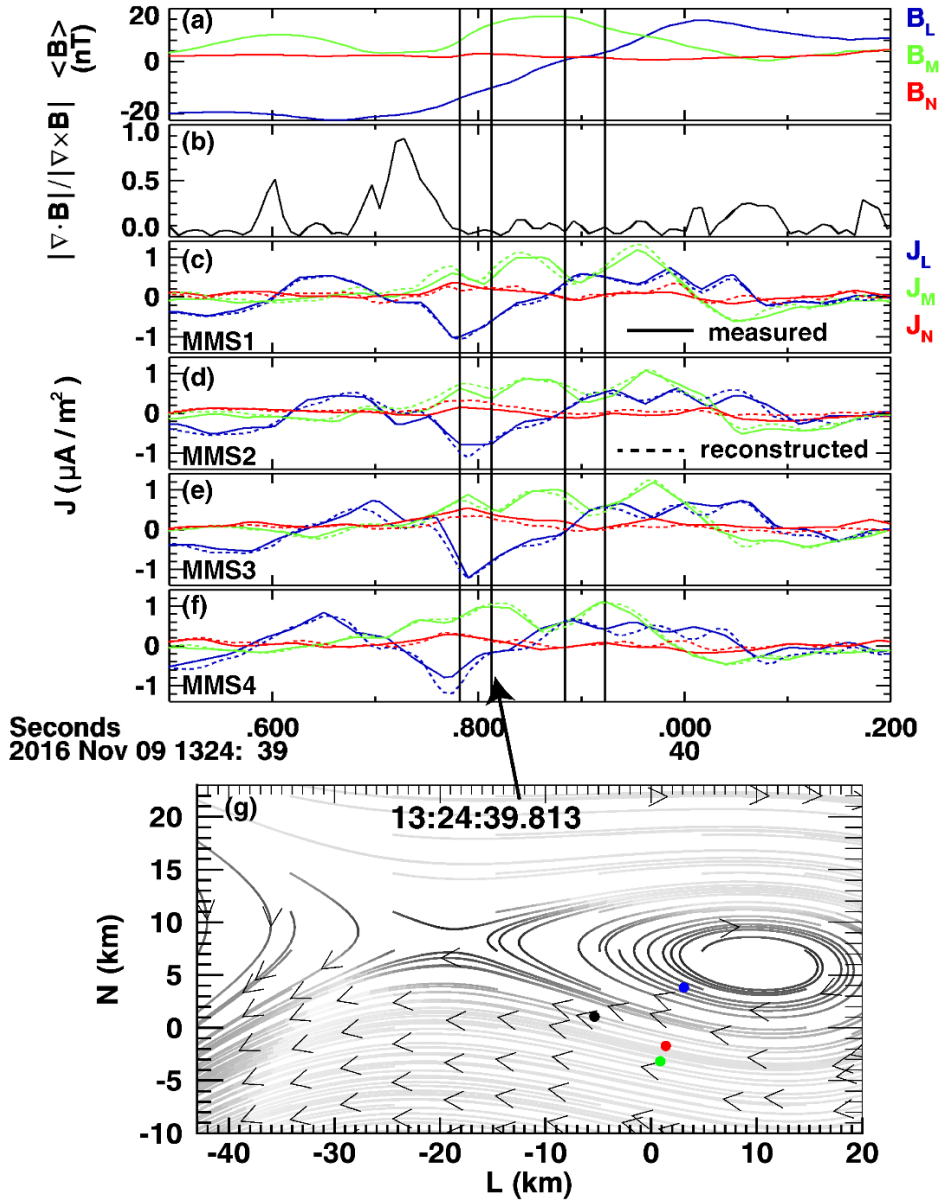

**Figure 3.** Reconstruction of the current sheet magnetic field using reduced 2$^{nd}$-order polynomial expansion. (a) magnetic field averaged over four spacecraft. In the two marked intervals, small values of $|\nabla \cdot B|/|\nabla \times B|$ (b), and the agreement between measured (solid) and reconstructed (dashed) current densities (c-f) for MMS1-4 serve as support of valid reconstruction. The reconstructed magnetic fields in the L-N plane at 13:24:39.813 UT is shown in (g), where an X-line exists at about 20 km away from the spacecraft. X-line exists in reconstructed fields during the two intervals marked by the vertical lines in (c)-(f).

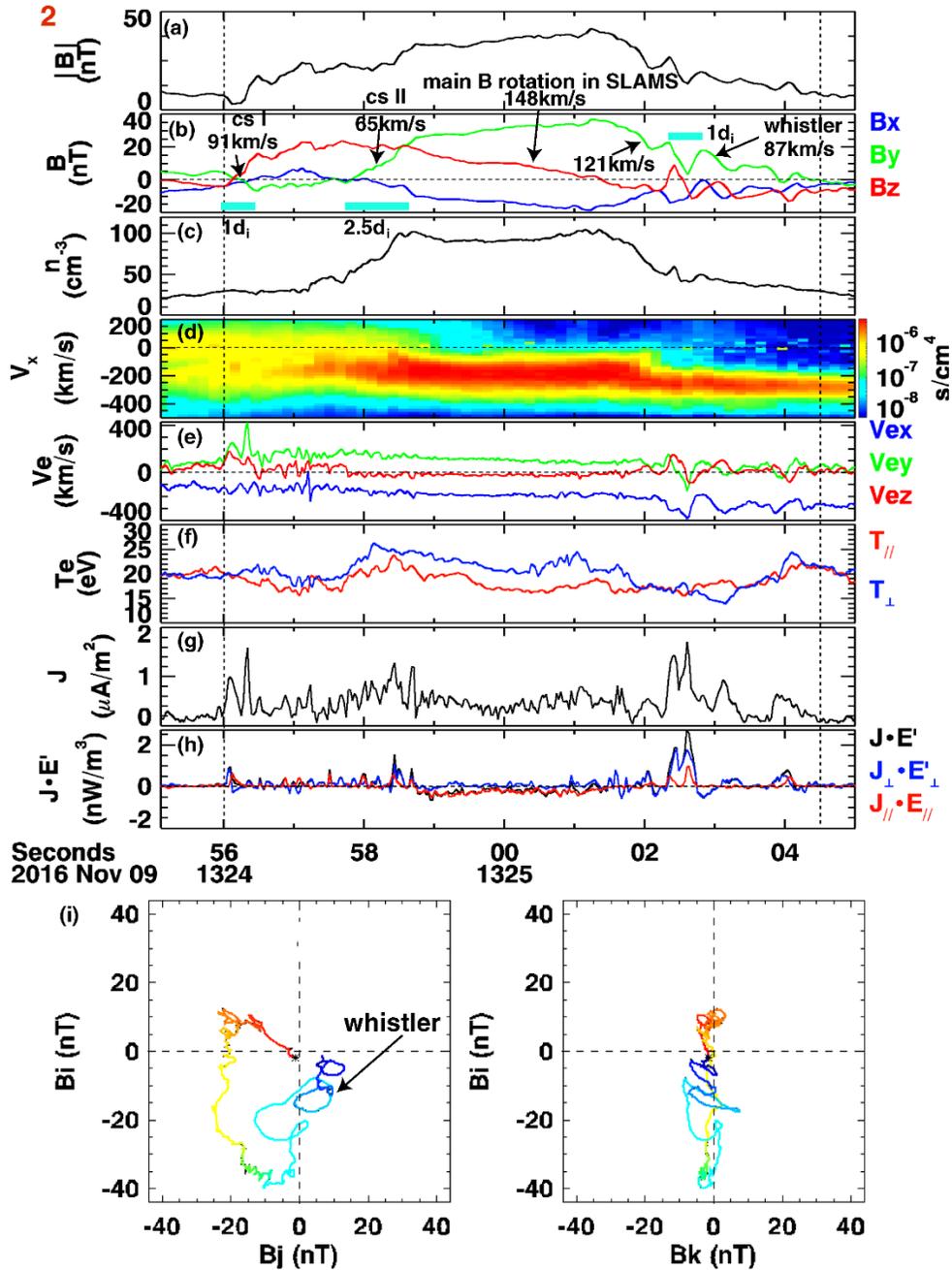

**Figure 4.** SLAMS event 2. Formats are the same as in Figures 1b-1j. The $d_i$-scale current sheets (cs I, cs II) as part of the magnetic field rotation near the downstream edge propagate much slower than the overall SLAMS, suggesting compression. Magnetosonic whistler waves with the wavelength of 1 $d_i$ exist at the upstream edge. Both lead to current density and energy conversion enhancements.

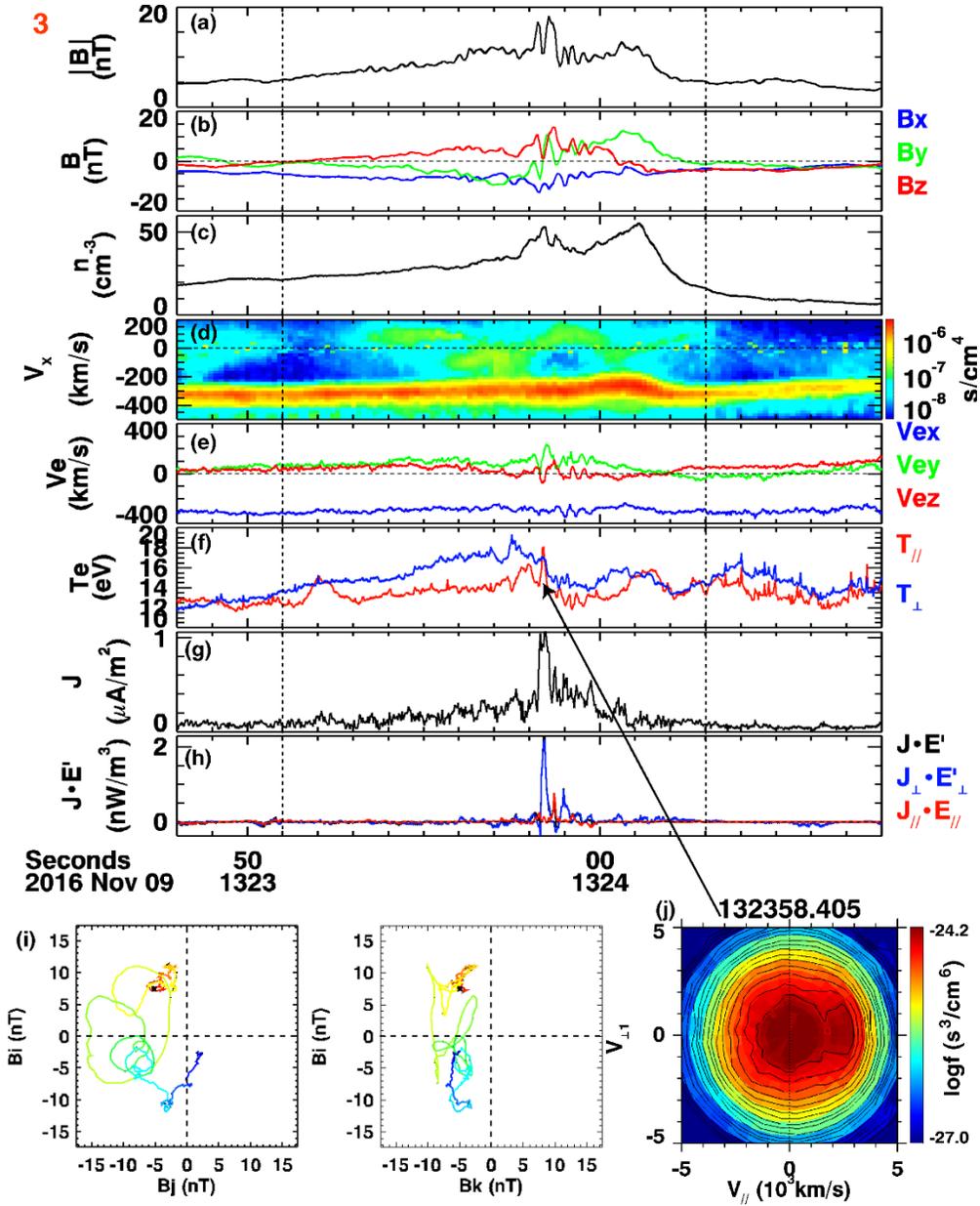

**Figure 5.** SLAMS event 3. (a)-(h) have the same formats as in Figures 1b-1i. The magnetosonic whistler waves in the SLAMS produce $d_i$-scale current density enhancements, and localized electron heating associated with a parallel electron beam (i). The overall perpendicular electron heating is associated with the magnetic field strength enhancement toward the center of the SLAMS, comparable to the localized parallel heating.

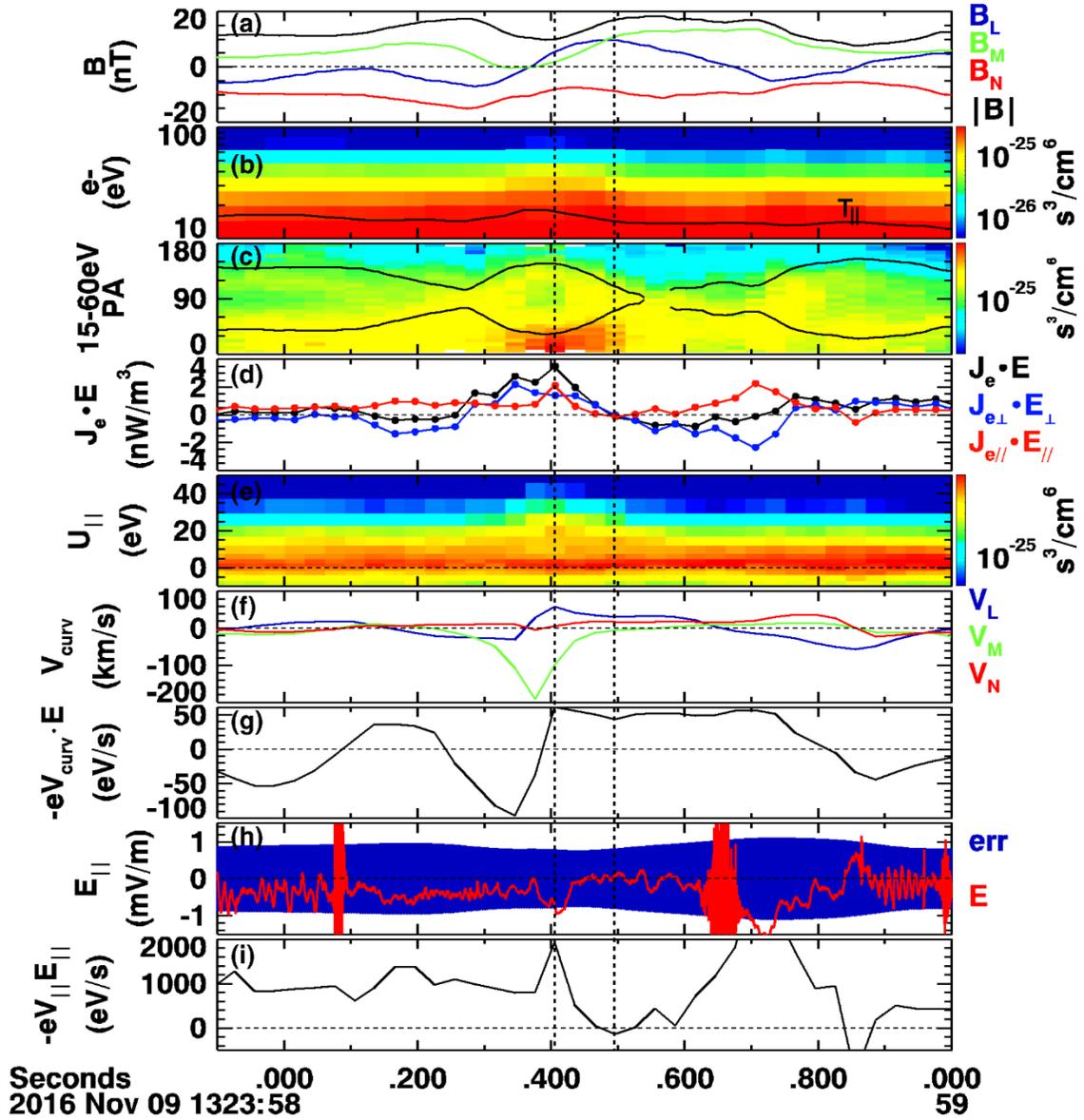

**Figure 6.** The current sheet in SLAMS event 3 with clear electron parallel energization. (a) Magnetic field in LMN, where L=[-0.240, 0.967, -0.081], M=[-0.459, -0.039, 0.888], N=[0.855, 0.251, 0.453] GSE. (b) Omni-directional electron spectrogram, where $T_{e\parallel}$ is overplotted. (c) Pitch angle distribution of 15-60 eV electrons. The black curves are the pitch angles for electrons that have PA=90° at |B|=17 nT near the current sheet edge. (d) Electron energy conversion rate $J_e \cdot E$, both are in the current sheet frame. (e) Electron distributions along the parallel energy. The distribution at the first vertical dashed line is energized by one bin (~3eV) compared to that at the second vertical dashed line. (f) Electron curvature drift velocity. (g) Energy conversion rate due to the curvature drift. (h) parallel electric field and its uncertainty. (i) Energy conversion rate due to the parallel electric field. (f)-(g) are evaluated at the barycenter of four spacecraft, while other panels are from MMS1.